# Nano-scale oxygen octahedral tilting in $0.90(Bi_{1/2}Na_{1/2})TiO_3$-$0.05(Bi_{1/2}K_{1/2})TiO_3$-$0.05BaTiO_3$ lead-free perovskite piezoelectric ceramics

Cheuk W. Tai[*][†] and Y. Lereah

Department of Physical Electronics, School of Electrical Engineering,

The Iby and Aladar Fleischman Faculty of Engineering,

Tel Aviv University, Ramat Aviv 69978, Israel

**Abstract**

The oxygen octahedral tilted domains in $0.90(Bi_{1/2}Na_{1/2})TiO_3$-$0.5(Bi_{1/2}K_{1/2})TiO_3$-$0.5BaTiO_3$ lead-free perovskite piezoelectric ceramic have been studied by transmission electron microscopy (TEM). Selected-area electron diffraction patterns shows the ½$ooo$ and ½$ooe$ reflections, indicating the presence of antiphase ($a^-a^-a^-$) and in-phase ($a^oa^oc^+$) octahedral tilting, respectively. The morphology and distributions of these tilted domains are shown in the centered dark-field images. Further, the Bragg-filtered high-resolution TEM image reveals that the size of the in-phase tilted domains varies from 1 to 8 nm across. The ceramic contains the mixture of non-tilted and variants of the antiphase and in-phase tilted domains.

---

[*] Electronic mail: cheukw.tai@gmail.com
[†] Present address: Department of Physical, Inorganic and Structural Chemistry, Arrhenius Laboratory, Stockholm University, S-106 91 Stockholm, Sweden.

**Introduction**

Environmental friendly piezoelectric materials have drawn great attention in recent years. $(Bi_{1/2}Na_{1/2})TiO_3$-based (abbreviated as BNT) perovskite solid solutions have been considered to be one of the most promising candidates. Analogous to the well-known perovskite piezoelectric material $Pb(Zr,Ti)O_3$,[1] the significantly improved piezoelectric and dielectric properties can be obtained in the compositions near the morphotropic phase boundary (MPB), at which both rhombohedral ($F_R$) and tetragonal ($F_T$) ferroelectric phases exist simultaneously. BNT has a rhombohedral perovskite structure (*R3c*) below its $T_c$ of 320 °C.[2,3,4,5] A number of solid solutions have been developed by adding the materials having the $F_T$ phase, such as $BaTiO_3$ (BT)[6] and $(Bi_{1/2}K_{1/2})TiO_3$ (BKT) (*P4mm*).[7] BNT-BKT and BNT-BT solid solutions show the improved dielectric and piezoelectric properties and easier electrically poling process. In addition, these compositions have relatively high depolarization temperature ($T_d$), at which the polarization and hence piezoelectricity vanish. In recent, a ternary system, BNT-BKT-BT, has been developed by mixing the above two solid solutions. It is shown that the dielectric and piezoelectric properties of some compositions near MPB are comparable with some $Pb(Zr,Ti)O_3$-based materials.[8,9,10]

Several studies have already reported that the phase transitions of BNT-based materials do not only relate to the cation displacements but also associate with oxygen octahedral tilting.[5,11,12,13,14,15,16] The antiphase octahedral tilting ($a^-a^-a^-$ in Glazer notation[17]) in BNT is found by the room temperature neutron diffraction[5,18] and transmission electron microscopy (TEM) studies.[13] An additional tetragonal phase with in-phase tilting ($a^o a^o c^+$) has been revealed by electron diffraction.[13] The coexistence of the $a^- a^- a^-$ and $a^o a^o c^+$ tilting has also been reported in Li-doped BNT-BKT-BT.[16] However, the $a^o a^o c^+$ tilting is only observed at the high-temperature phase of Na-rich BNT-BKT[5] and BNT ceramics studied by neutron diffraction.[5,19]

In this study, the octahedral tilting in 0.90BNT-0.5BKT-0.5BT ceramic is characterized by electron diffraction, centered dark-field (CDF) and high-resolution transmission electron microscopy (HRTEM) imaging. The ceramic is an optimized composition in the ternary solid solution in order to have the higher piezoelectric constant and remanent polarization and the lower coercive field than those in BNT-BKT and BNT-BT and also a reasonable high $T_d$.[9,10] 0.90BNT-0.5BKT-0.5BT ceramics is close to MPB but the room-temperature x-ray diffraction studies showed that the macroscopic crystal structure is rhombohedral ($F_R$).[9,20] However, neither type of oxygen octahedral tilting has been reported.

**Experimental**

0.90BNT-0.5BKT-0.5BT ceramics were prepared by conventional mixed oxide route.[9] For the TEM study, samples were mechanically polished to a thickness of about 20 μm. Ion-milling was then performed to obtain electron-transparency sample. A 200 kV field-emission electron microscope (FEI Tencai F20ST) was used to record the TEM results at room temperature. For simplicity, the space group *Pm3m* is used to index the electron diffraction patterns.

**Observations and discussion**

Different zone axes selected-area electron diffraction (SAED) patterns of the ceramics are shown in Figure 1. The region of interest is selected by an aperture, of which the diameter is 200 nm. Figs. 1(a) and (b) are the SAED patterns taken along a <100> and <111> zone axis, respectively, in which the ½*ooe* (where 'o' and 'e' indicate an index with odd and even number, respectively) reflections are observed besides the fundamental perovskite reflections. This indicates the presence of the in-phase octahedral tilting ($a^o a^o c^+$). It should notice that in $a^o a^o c^+$ tilting the condition for the allowed additional reflections is $h \neq \pm k$. However, The 1/2 1/2 0 reflection, which is kinematically forbidden, often appears in SAED pattern because of the occurrence of double diffraction such as the route 3/2 3/2 0 + -1 -1 0.[13,14,15,16] This tilt system that occurred at room temperature has only been confirmed by the TEM studies,[13,16] indicating its short-range and disorder nature. We observe that the intensity of the ½*ooe* reflections is stronger than that in the previous TEM study of BNT.[13] It suggests that the doped $K^+$ and $Ba^{2+}$ stabilize the tetragonal phase at room temperature and hence the $a^o a^o c^+$ tilting. A SAED pattern taken along a <110> zone axis is shown in Fig. 1(c), in which the ½*ooe* and ½*ooo* reflections are observed simultaneously. It is known that in the $F_R$-phase of BNT-based solid solutions the ½*ooo* reflections is contributed by the antiphase octahedral tilting ($a^- a^- a^-$) instead of the A-site cation ordering.[5,13,19] Similar to $a^o a^o c^+$ tilting, the kinematically forbidden 1/2 1/2 1/2 reflection is given by double diffraction although the condition for the allowed additional reflections is $h \neq \pm k; k \neq \pm l; l \neq \pm h$. However, we confirmed that neither one of the four mixed tilted system exists in the ceramic because the ½*oee* reflections has never been observed, similar to in other BNT-based materials. It should be noted that not all <110> zone axes show the forbidden reflections because some reflections are absence when equal magnitude tilting occurs about different orthogonal axes.[21] Such observation has been reported in the TEM study of Li-doped BNT-BKT-BT ceramics.[16]

Figure 2(a) shows the microscopic ferroelectric domains viewed along a direction close to [111] zone axis. A relatively large domain wall is shown by the dark contrast. The wall is not as sharp as in the typical $F_R$-phase ferroelectrics. A number of nano-sized domains are across the wall. The orientation of this domain wall does not follow any crystallographic plane that is similar to that in some $F_R$-phases of PZT.[22,23] In addition, some tweed domains, of which the width is less than 10 nm, are observed in the microscopic ferroelectric domain. It is known that tweed domains exist in many MPB solid solutions.[23,24] Some tweed domains walls are parallel to (01-1), which is marked by a white dashed line. Figure 2b is a CDF image of a ½*ooe* reflection in a region selected in Fig. 2a. It can be seen that the outline of the microscopic ferroelectric domains can be roughly traced by the distribution of the in-phase tilted domains (bright contrast). However, we do not see a direct relationship between the in-phase tilted domains and the ferroelectric domains. It should be noticed that the dark contrast represents the corresponding region having either the non-tilted crystal structure or in-phase tilted domains having a different tilting axis. The inset in Fig. 2(b) clearly shows that the clusters are composed of the separated nano-scale tilted domains.

Fig. 2(c) shows the ferroelectric domains viewed close to the [0-11] zone axis. The width of the microscopic parallel-band domains is from 10 to 20 nm. The domain walls are at the inclined (-110) or (101). The termination of the domain walls does not follow in any crystallographic plane. These observations are similar to those in $F_R$-phase PZT ceramics near MPB.[22] Figure 2(d) is a CDF image of a ½$ooo$ reflection in the selected region, which is indicated by a white circle in Fig. 2(c). Similar to Fig. 2(b), the morphology of some ferroelectric domains can be outlined by the occupancy of the antiphase tilted domains shown in the CDF image (Fig. 2(d)). In contrast to the in-phase tilting, $a^-a^-a^-$ tilting is more apparent because it has a higher volume fraction in this ceramic at room temperature, as well as in BNT.[5,13] It is found that a number of ferroelectric domains are mostly filled by the antiphase tilted domains.

Figure 3(a) is a HRTEM image showing three sets of {110} lattice fringes along a <111> zone axis. The intensity of the additional modulations of the multiple of {110} given by in-phase octahedral tilting is difficult to be seen, but the corresponding reflections are observed in the diffractogram of the HRTEM image (see Fig. 3(b)). In contrast to the diffraction pattern shown in Fig. 1(b), the three-fold symmetry is not found in this Fourier transformed pattern. However, it shows one of the three different variants of the in-phase $a^o a^o c^+$ tilting. In order to characterize the individual nano-sized tilted domain, Bragg-filtering is applied to the HRTEM image. Fig. 3(c) is the filtered HRTEM image showing the lattice fringes corresponding to a ½$ooe$ reflection, which are marked by the white arrows in Fig. 3(b). It is found that the size of the in-phase tilted domain varies from 1 to 8 nm across. The dark contrast regions in the filtered image represent the non-tilted region and possibly the antiphase tilted domains, which are unable to be observed at this zone axis.

## Summary

We conclude that no macroscopic octahedral tilted domain exists and the ceramic contains the mixture of non-tilted regions, nano-sized in-phase ($a^o a^o c^+$) and antiphase ($a^- a^- a^-$) octahedral tilted domains with their variants in different orientations. The origin of the nano-sized octahedral tilted domains is highly likely related to the structural disorder of Bi atoms that gives rise to the local structure and deviates from the macroscopic average one.[11] These nano-sized tilted domains influence the configuration of the ferroelectric (or ferroelastic) domains and their walls by the strains induced at the interfaces between different kinds of the tilted domains and their variants. It is deduced that the local structures and localized strains break the long-range ferroelectric order. We suggest that the frequency dispersion in the temperature-dependent dielectric curves is contributed by not only the change of polar phase but also the nano-scale oxygen octahedral titling, which are connected to the local chemistry in the solid solution.

## Acknowledgement

We are grateful to Dr. S. H. Choy (The Hong Kong Polytechnic University) for providing the quality samples. This study was supported by European Commission under Grant No. 29637.

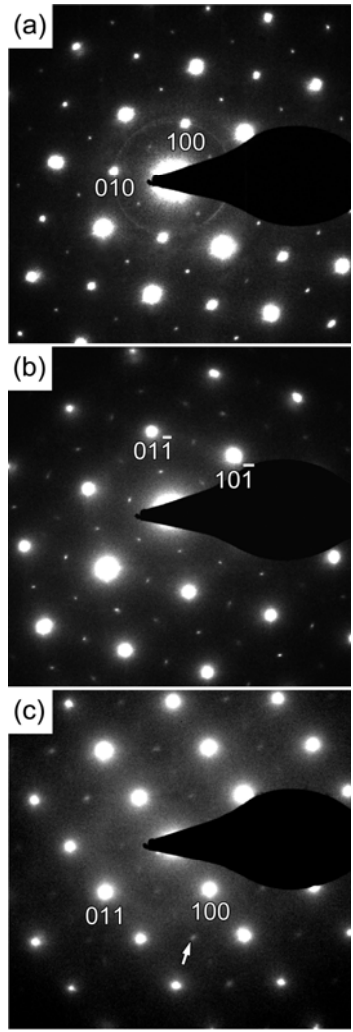

FIG 1. Selected-area electron diffraction patterns taken along a (a) <100>, (b) <111> and (c) <110> zone axis. The ½*ooe* reflections exist in all three diffraction patterns and the ½*ooo* reflections (indicated by an arrow) appear in the <110> zone axis only.

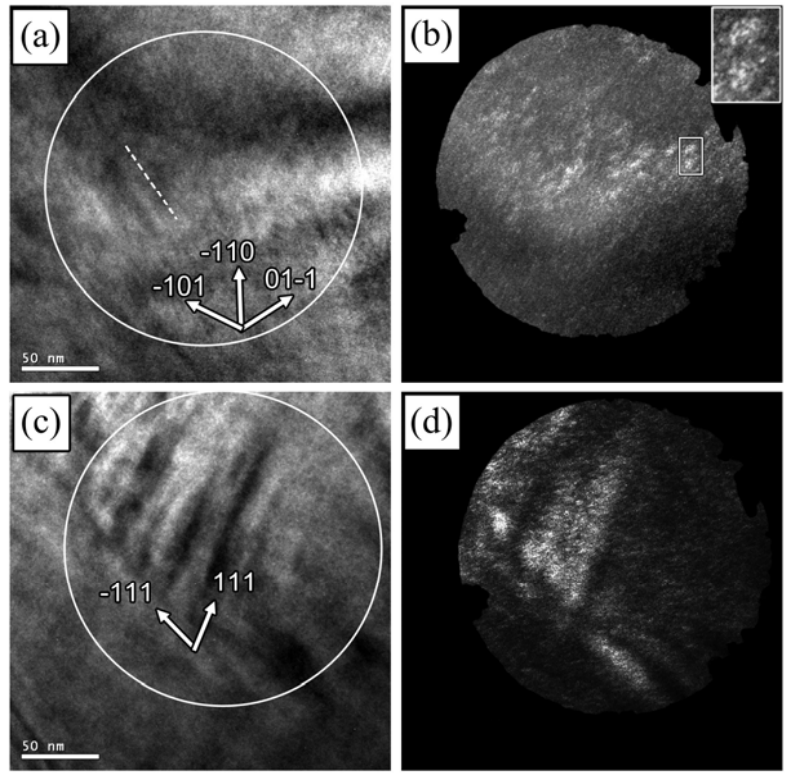

FIG 2. The microscopic ferroelectric domains viewing along the beam direction close to (a) [111] and (c) [0-11] zone axis, respectively. The white dashed line shown in (a) is the trace of the (01-1). Centered dark-field images of the (b) ½ooe and (d) ½ooo reflection in the selected region (indicated by a white circle) shown in (a) and (c), respectively. Inset in (b): an enlarge image showing the clusters of in-phase tilted domains.

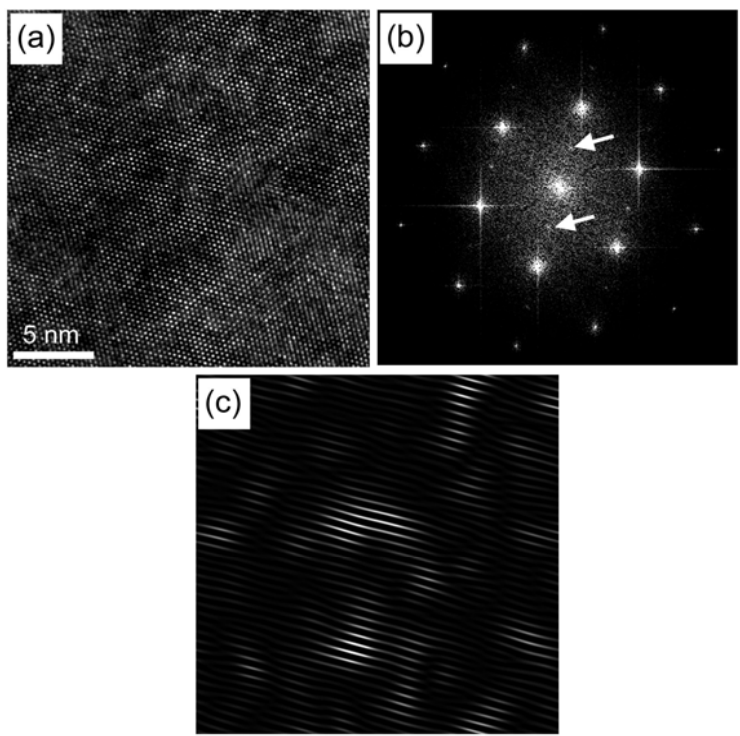

FIG 3. (a) HRTEM image of the ceramic taken along a <111> zone axis. (b) Fourier transform of the HRTEM image. The reflections ½ooe are indicated by white arrows. (c)Bragg-filtered image of (a) shows the morphology of the nano-sized in-phase tilted domains.